\documentclass[onecolumn,pre]{revtex4}
\usepackage{bm}
\usepackage{epsfig}
\usepackage{amsmath}
\begin{document}
\title{Electrochemical electron transfer: An Analytically Solvable Model}
\author{Aniruddha Chakraborty \\
School of Basic Sciences, Indian Institute of Technology Mandi,\\
Mandi, Himachal Pradesh, 175001, India}
\date{\today }
\begin{abstract}
\noindent  We propose an analytical model based on diffusion-reaction equation approach for electrochemical electron transfer reaction, where the rate is limited by the electron transfer process. The electron transfer from an ion in solution to the metal electrode would occur as soon as the energy of the orbital on the ion matches the Fermi energy of the metal and a new ion with more positive charge is formed. Obviously the ion before elctron transfer and the new ion, which is formed after electron has transferred, moves under the influence of different potentials. The coupling between these two potentials is assumed to be represented by a Dirac Delta function. The diffusive motion in this paper is described by the Smoluchowskii equation. Our solution requires only the knowledge of the Laplace transform of the Green's function for the motion in both the uncoupled potentials. Our model is more general than all the earlier models, beacuse we are the first one to consider the potential of both the ions explicitely.
\end{abstract}
\maketitle
The theory of electron transfer reactions in solution is an important topic \cite{Marcus}. It is conventional to use transition state theoretical (TST)
approach to the problem. TST provides the rate expression as $k = A e^{- \Delta G/RT}$, where $\Delta G$ is the free energy of activation and $A$ is the pre-exponential factor. The TST has the limitation of not taking account of the deviation of the reactant population from the equilibrium one \cite{Zusman,Calef,Wolynes,Hynes,Rips}. For fast reaction (in comparison with the rate of attainment of thermal equilibrium ) the deviation from equilibrium population can be very important. Homogeneous electron transfer is one of such cases and a diffusion-reaction equation approach is suggested for it \cite{Sumi,Nadler}. In the
electrochemical case, magnitude of reaction rate is very sensitive to overpotential and so it is most likely that there are regimes where the deviation from equilibrium population becomes significant. In the following, we investigate the problem of electrochemical electron transfer using a diffusion-reaction approach. The case where electron transfer is the rate determining step, rather than diffusion from the bulk, is considered here. We use Smoluchowski equation to understand the reaction dynamics. We start with an ion, in a polar solvent, near the surface of a metal electrode. It has an occupied orbital $\left| a \right >$ in it, which may give an electron to the electrode. This orbital interacts with the electrons of the metal and as a result of which it is broadened. The energy of the orbital $\left| a \right >$ depends on the extent of polarization of the surrounding solvent and its value, when the surrounding polarization is in thermal equilibrium as assumed to be $\epsilon_a$. Motion of the surrounding solvent will obviously change the energy of the orbital and we denote this change by $Q$. So at any instant of time $t$, the energy of the orbital may be written as $\epsilon_a+Q(t)$. It is fairly easy to derive the corresponding Fokker-Planck equation for the probability $P_1(Q,t)$ that the shift has a value $Q$ at the time $t$ to be
\begin{equation}
\frac{\partial P_1(Q,t)}{\partial t} = \frac{2kT \lambda}{\tau_L} \frac{\partial^2 P_1(Q,t)}{\partial Q^2}+\frac{1}{\tau_L}\frac{\partial}{\partial Q}P_1(Q,t),
\end{equation}
where $\tau_L$is the time of longitudinal relaxation and $\lambda$ is the solvent re-organization energy. If $\epsilon_a+ Q(t)$ moves above the Fermi energy $\epsilon_F$ of the metal, then an electron can be transferred to the electrode. This implies that we are considering a metal of a fairly large width in energy and also we assume the rate of electron transfer
is independent of the energy of the orbital $\left| a \right >$, as long as it is above the Fermi level of the metal. Therefore, the probability $P_1(Q,t)$ is now given by the following equation \cite{Chakravarti}
\begin{equation}
\frac{\partial P_1(Q,t)}{\partial t} = \frac{2kT \lambda}{\tau_L} \frac{\partial^2 P_1(Q,t)}{\partial Q^2}+\frac{1}{\tau_L}\frac{\partial}{\partial Q}P_1(Q,t)- K(Q)P_1(Q,t),
\end{equation}
In the above equation has a sink term, $K(Q)$. $Q_F= \epsilon_F- \epsilon_a$ is that particular value of $Q$, at which the energy of the orbital is equal to the Fermi energy $\epsilon_F$. $P_1(Q,t)$ is now the probability that the shift has a value $Q$ at the time $t$, and that the electron has not been transferred to the electrode. So $P_1(Q,t)$ is actually the probability that an ion, placed in the reaction zone at the time $t = 0$, would still survive as the same ion at time $t$. The Eq.(2) is solved by Sebastian {\it et. al.} \cite{Chakravarti,Sebastian} only under steady state approximation. In our model we consider the survival probability of the new ion (formed after the electron is transferred to the electrode), which is denoted by $P_2(Q,t)$. So in our model the relevant equation is
\begin{eqnarray}
\frac{\partial P_1(Q,t)}{\partial t} = \frac{2kT \lambda_1}{\tau_{L_1}} \frac{\partial^2 P_1(Q,t)}{\partial Q^2}+\frac{1}{\tau_{L_1}}\frac{\partial}{\partial Q}P_1(Q,t)- K(Q)P_2(Q,t)\\ \nonumber
\frac{\partial P_2(Q,t)}{\partial t} = \frac{2kT \lambda_2}{\tau_{L_2}} \frac{\partial^2 P_2(Q,t)}{\partial Q^2}+\frac{1}{\tau_{L_2}}\frac{\partial}{\partial Q}P_2(Q,t)- K(Q)P_1(Q,t)
\end{eqnarray}
In the above equation has a sink term, which we assume to be represented by a Dirac Delta function, $K(Q) = k_0\delta(Q-Q_F)$.
The survival probability $P_1(t)$ is defined using $P_1(Q,t)$ by averaging over the entire range of $Q$.
\begin{equation}
P_1(t)=\int dQ P_{1}(Q,t).
\end{equation}
Eq.(3) can be re-written as given below
\begin{eqnarray}
\frac{\partial P_1(Q,t)}{\partial t} = {\cal L}_1 P_1(Q,t) - k_0\delta(Q-Q_F)\delta(Q-Q_F) P_2(Q,t)  \\ \nonumber
\frac{\partial P_2(Q,t)}{\partial t} ={\cal L}_2 P_2(Q,t) - k_0\delta(Q-Q_F) P_1(Q,t). \nonumber
\end{eqnarray}
In the above 
\begin{equation}
{\cal L}_i= \frac{2kT \lambda_i}{\tau_{L_i}} \frac{\partial^2 }{\partial Q^2}+\frac{1}{\tau_{L_i}}\frac{\partial}{\partial Q}.
\end{equation}
In the follwoing we provide a general procedure for finding the exact analytical solution of Eq. (5). The Laplace transform ${\cal P}_i(Q,s)=\int_{0}^{\infty} P_i(Q,t)e^{-st} dt$ obeys
\begin{eqnarray}
[s-{\cal L}_1] {\cal P}_1(Q,s)+k_0 \delta(Q-Q_F) {\cal P}_2 (Q,s) = P^0_1(Q_0) \\ \nonumber
[s-{\cal L}_2] {\cal P}_2(Q,s)+k_0 \delta(Q-Q_F) {\cal P}_1 (Q,s) = 0, \nonumber
\end{eqnarray}
where $P^0_1(Q_0)=P_1(Q,0)$ and $P_2(Q,0)=0$ are the initial distributions. 
\begin{equation}
 \left(
\begin{array}{c}
{\cal P}_1 (Q,s) \\
{\cal P}_2 (Q,s)
\end{array}
\right) = \left(
\begin{array}{cc}
s-{\cal L}_1 & k_0 \delta(Q-Q_F) \\
k_0 \delta(Q-Q_F) & s-{\cal L}_2
\end{array}
\right)^{-1}
\left(
\begin{array}{c}
P^0_1(Q) \\
0
\end{array}
\right)  ,
\end{equation}
Using the partition technique \cite{Lowdin}, solution of this equation can be expressed as 
\begin{equation}
{\cal P}_1(Q,s)=\int_{-\infty}^{\infty} dQ_0 G(Q,s;Q_0)P^0_1(Q_0),
\end{equation}
where $G(Q,s;Q_0)$ is the Green's function defined by
\begin{equation}
G(Q,s;Q_0)=\left < Q \left|[s-{\cal L}_1 - {k_0}^2 S[s-{\cal L}_2]^{-1}S]^{-1}\right| Q_0 \right>
\end{equation}
After simplification
\begin{equation}
G(Q,s;Q_0)=\left < Q \left|[s-{\cal L}_1 - {k_0}^2 G^{0}_2(Q_F,s;Q_F) S ]^{-1}\right| Q_0 \right>,
\end{equation}
where
\begin{equation}
G^{0}_2(Q,s;Q_0)=\left < Q \left|[s-{\cal L}_2 ]^{-1}\right| Q_0 \right>
\end{equation}
Now we use the operator identity
\begin{equation}
[s-{\cal L}_1  - {k_0}^2 G^{0}_2(Q_F,s;Q_F) S]^{-1}=[s-{\cal L}_1]^{-1}+[s-{\cal L}_1]^{-1}{k_0}^2 G^{0}_2(Q_F,s;Q_F) S [s-{\cal L}_2  - {k_0}^2 G^{0}_2(Q_F,s;Q_F) S]^{-1}
\end{equation}
Inserting the resolution of identity $I=\int_{-\infty}^{\infty} dy \left|y \left> \right < y \right|$ in the second term of the above equation and integrating, we arrive at an equation which is similar to Lippman-Schwinger equation.
\begin{equation}
G(Q,s;Q_0)=G^0_1(Q,s;Q_0) + {k_0}^2 G^0_1(Q,s;Q_F)G^0_2(Q_F,s;Q_F)G(Q_F,s;Q_0).
\end{equation}
where $G^0_1(Q,s;Q_0)=\left < x \left|[s-{\cal L}_1]^{-1}\right| Q_0 \right>$. We now put $Q=Q_F$ in the above equation and solve for $G(Q_F,s;Q_0)$ to get
\begin{equation}
G(Q,s;Q_0)=\frac{G^0_1(Q,s;Q_0)}{1- {k_0}^2 G^0_1(Q_F,s;Q_F)G^0_2(Q_F,s;Q_F)}.
\end{equation}
This when substitued back into Eq. (14) gives
\begin{equation}
G(Q,s;Q_0)=G^0_1(Q,s;Q_0) + \frac{{k_0}^2 G^0_1(Q,s;Q_F)G^0_2(Q_F,s;Q_F)G^0_1(Q_F,s;Q_0)}{1-{k_0}^2 G^0_1(Q_F,s;Q_F)G^0_2(Q_F,s;Q_F)}.
\end{equation}
Using this Green's function in Eq. (9) one can caluclate ${\cal P}_1(Q,s)$ explicitely. Here we are interested to know the survival probability $P_1(t)$, which is given by
\begin{equation}
P_1(t) = \int_\infty^\infty dQ P(Q,t).
\end{equation}
It is possible to evaluate Laplace Transform  ${\cal P}_1(s)$ of $P_1(t)$ directly. ${\cal P}_1 (s)$ is defined in terms of ${\cal P}(Q,s)$ by the following equation,
\begin{equation}
{\cal P}_1(s)=\left(1+\left[1-k_0^2 G^0_1(Q,s;Q_F)G^0_2(Q_F,s;Q_F)\right]^{-1}k_0^2 G^0_2(Q_F,s;Q_F)\int^{\infty}_{-\infty}dQ_0 G^0_1(Q_F,s;Q_0)P^0_1(Q_0)\right)/(s).
\end{equation}
From the above equation we see that ${\cal P}_1(s)$ depends on $G^0_2(Q_F,s;Q_F)$ which is different from the results of all earlier studies \cite{Chakravarti,Sebastian}. The average and long time rate constants can be found from ${\cal P}_1(s)$. Thus, $k^{-1}_{1}={\cal P}_1(0)$ and $k_{L}= - ($ pole of $\left[1-k_0^2 G^0_1(Q,s;Q_F)G^0_2(Q_F,s;Q_F)(s)\right]^{-1})$, closest to the origin, on the negative $s$ - axis, and is independent of the initial distribution but depends on $G^0_2(Q_F,s;Q_F)$. The expression that we have obtained for ${\cal P}_1(s)$, $k_I$ and $k_L$ are quite general and are valid for any set of potentials.

\end{document}